\documentstyle[emulateapj,psfig,lscape]{article}

\lefthead{HU, COWIE, \& M{\sc{c}}MAHON}
\righthead{DENSITY OF Ly$\alpha$ EMITTERS AT VERY HIGH REDSHIFT}
\newcommand{\civ}{{\rm C}\thinspace{\sc{iv}}}

\newcommand{\nii}{[N\thinspace{\sc{ii}}]}
\newcommand{\oii}{[O\thinspace{\sc{ii}}]}
\newcommand{\oiii}{[O\thinspace{\sc{iii}}]}
\slugcomment{To appear in The Astrophysical Journal (Letters)}

\begin{document}
\title{The Density of Ly{$\boldmath\alpha$} Emitters at Very High Redshift}
\author{Esther M. Hu\altaffilmark{1} and Lennox L. Cowie\altaffilmark{1}}
\affil{Institute for Astronomy, University of Hawaii, 2680 Woodlawn
  Drive, Honolulu, HI 96822\\
  hu@ifa.hawaii.edu, cowie@ifa.hawaii.edu} 
\and
\author{Richard G. McMahon\altaffilmark{1}}
\affil{Institute of Astronomy, University of Cambridge, Madingley Road,
  Cambridge CB3\thinspace{0HA}\\
  rgm@ast.cam.ac.uk}
\altaffiltext{1}{Visiting Astronomer, W. M. Keck Observatory,
  jointly operated by the California Institute of Technology, the University
  of California, and the National Aeronautics and Space Administration.}

\begin{abstract}
We describe narrowband and spectroscopic searches for emission-line
star forming galaxies in the redshift range 3 -- 6 with the 10 m Keck\,II
Telescope. These searches yield a substantial population of
objects with only a single strong (equivalent width $\gg 100$ \AA)
emission line, lying in the $4000 - 8500$ \AA\ range.  Spectra of the
objects found in narrowband--selected samples at $\lambda \sim5390$ \AA\ and
$\sim6741$ \AA\ show that these very high equivalent width emission
lines are generally redshifted Ly$\alpha\ \lambda\,1216$ \AA\ at
$z\sim3.4$ and 4.5.  The density of these emitters above the 5$\sigma$
detection limit of $1.5\times 10^{-17}$ ergs cm$^{-2}$ s$^{-1}$ is
roughly 15,000/$\sq^{\circ}$/unit $z$ at both $z\sim3.4$ and 4.5.  A
complementary deeper ($1\ \sigma \sim 10^{-18}$ ergs cm$^{-2}$ s$^{-1}$)
slit spectroscopic search covering a wide redshift range but a more
limited spatial area ($200\ \sq''$) shows such objects can be found
over the redshift range $z=3 - 6$, with the currently highest redshift
detected being at $z=5.64$.  The Ly$\alpha$ flux distribution can be
used to estimate a minimum star formation rate in the absence of
reddening of roughly $0.01\ M_{\odot}$ Mpc$^{-3}$ yr$^{-1}$ ($H_0 =
65\ {\rm km}\ {\rm s}^{-1}\ {\rm Mpc}^{-1}$, $q_0 = 0.5$).  Corrections
for reddening are likely to be no larger than a factor of two, since
observed equivalent widths are close to the maximum values obtainable from
ionization by a massive star population. Within the still significant
uncertainties, the star formation rate from the Ly$\alpha$--selected
sample is comparable to that of the color-break--selected samples at
$z\sim 3$, but may represent an increasing fraction of the total rates at
higher redshifts.  This higher-$z$ population can be readily studied with
large ground-based telescopes.
\end{abstract}


\section{Introduction}

The search for high-redshift galaxies and the effort to map the star
formation history of galaxies have progressed rapidly in the last
several years as magnitude--limited spectroscopic surveys 
pushed into the $z=1-3$ range (Cowie et al.\markcite{large_sample}
1996; Cohen et al.\markcite{cohenhdf} 1996), while color--based selection
techniques produced many objects in the $z=2.5-5$
range (Steidel et al.\markcite{stei96a} 1996a, \markcite{stei96b}1996b;
Franx et al.\markcite{franx} 1997; Steidel et al.\markcite{stei98}
1998), particularly in the exquisite Hubble Deep Field (HDF) sample
(Lowenthal et al.\markcite{low97} 1997).  However, the galaxies chosen
by these techniques correspond to objects with ongoing massive star
formation and small amounts of extinction, and may represent only part
of the populations at these early epochs.  More evolved objects may be
heavily dust--reddened and more easily picked out at submillimeter
wavelengths (Smail, Ivison, \& Blaine\markcite{smail97} 1997; Hauser et
al.\markcite{dirbe} 1998), while earlier stages in evolution may
have relatively little continuum light and be too faint to be seen in
the magnitude--limited samples or selected with the color-break
techniques at current sensitivity limits.

This latter class of objects may represent the earliest stages of the
galaxy formation process, in which substantial amounts of metals have yet
to form.  These galaxies may have much stronger Ly$\alpha$ emission
relative to the stellar continuum, since they have massive star
formation that can excite the Ly$\alpha$ emission line, but without so
much dust that the line is suppressed, and this can result in very high
observed equivalent widths in the range of 100--200\,(1+$z$) \AA\ 
(e.g., Charlot \& Fall\markcite{charl93} 1993).  Such objects may
be hard to pick out with color-break techniques but be detectable in
Ly$\alpha$ searches of sufficient depth (Cowie\markcite{cowie88} 1988;
Thommes\markcite{thommes} 1996).  An increased incidence of
strong Ly$\alpha$ emission does, indeed, appear in the color-break samples 
at fainter continuum magnitudes (Steidel et al.\markcite{stei98} 1998). 

Earlier blank-field Ly$\alpha$ surveys (e.g., Thompson, Djorgovski, \&
Trauger\markcite{tdt95} 1995; Thompson \& Djorgovski\markcite{td95} 1995) 
failed to find such objects, as Pritchet\markcite{pri94} (1994) has 
summarized, but their sensitivity lay at the margin of where such 
objects would be expected in significant numbers (Cowie\markcite{cowie88} 
1988; Thommes \& Meisenheimer\markcite{thommes95} 1995).  However, Hu \& 
McMahon\markcite{br2237} (1996), using very deep targeted narrowband searches
($1\ \sigma = 1.5 \times 10^{-17}$ ergs cm$^{-2}$ s$^{-1}$), found $z\sim4.55$
Ly$\alpha$-emitting galaxies with the very strong emission and weak or
undetected continuua predicted for early star-forming objects. The
advent of 10 m telescopes, along with this successful detection of
Ly$\alpha$ emitters, prompted us to undertake a new survey that has
been successful in detecting blank-field high-$z$ Ly$\alpha$ emitters.
The present Letter describes the early results of this search, which
uses extremely deep narrowband filter exposures taken with LRIS
(Oke et al.\markcite{lris} 1995) on the Keck II telescope to
search for emission-line populations at extremely faint flux levels
($1\ \sigma=3 \times 10^{-18}$ ergs cm$^{-2}$ s$^{-1}$).  This survey picks
out sources of extremely high equivalent width ($W_{\lambda}>100$ \AA)
emission lines as candidates, and then uses followup LRIS spectroscopy
(\S2) to determine if these are Ly$\alpha$ emitters.

The first results from the Hawaii survey with a 5390/77 \AA\ filter,
corresponding to Ly$\alpha$ emission at $z\sim3.4$, were described in
Cowie \& Hu\markcite{smitty1} 1998 (hereafter, Paper I), and yielded a
number of candidate high-$z$ galaxies similar to the redshift $z\sim4.55$
Ly$\alpha$-emitting galaxies found by Hu \& McMahon\markcite{br2237}
(1996) and emitters at $z\sim 2.4$ (Pascarelle et
al.\markcite{pasc96} 1996; Francis\markcite{fra97} et al.\ 1997)
in targeted searches.  In Paper I, we showed
that the use of color selection on emission-line selected objects of high
equivalent width ($> 100$ \AA) picked out objects with continuum colors
similar to those of color--selected Lyman break galaxies with measured
redshifts of $z\sim3.4$.  They also recovered the one field object whose 
previously measured redshift placed Ly$\alpha$ within the filter
bandpass.  However, the emission-line galaxies selected by their high
equivalent width also comprised objects with very faint continuua, that
would have fallen below the magnitude threshold of current Lyman break
surveys, and also included two objects that could not be detected in
Keck imaging of the optical continuum ($1\ \sigma$ $B=27.8$, $V=27.5$,
and $I=25.8$; $W_{\lambda}>400$ \AA).

In the present Letter we first present spectroscopic followups for the
narrowband candidates of Paper I.  In a small fraction of the cases,
the spectrum shows both Ly$\alpha$ and \civ\ $\lambda$ 1550 \AA,
confirming the redshift identification but suggesting AGN-like
properties.  However, the majority of the spectra show only a single
strong emission line.  The absence of other detectable features,
in combination with the high
equivalent width of the selected candidates, identifes the single line as
redshifted Ly$\alpha$ emission, and argues that the equivalent
width criterion ($W_{\lambda}\gg 100$ \AA) can, in fact, be used as a
good diagnostic of high-$z$ Ly$\alpha$-emitting galaxies.  We then
(\S3) present results of a second deep narrowband search with a 78
\AA\ bandpass $\lambda \sim6741$ \AA\ filter (Ly$\alpha$ at $z\sim 4.54$) and
followup spectroscopy, that confirmed two Ly$\alpha$-emitting
galaxies at this higher redshift. In \S4, we describe a very deep
blank-field spectroscopic search (6 hr LRIS integration 
covering $\lambda\lambda\,\sim5000-10000$ \AA) that yielded four emitters at
redshifts 3.04 -- 5.64.  Finally (\S5), the data on emission-line
objects from the imaging surveys in the two redshift intervals are
combined with various spectroscopic surveys at lower redshift to show
the evolution of the emission-line fluxes with redshift.
We argue that the
Ly$\alpha$-emitting objects are significant contributors to the
integrated star formation of the galaxy population throughout the
$z=3 - 6$ redshift range, and that the integrated star formation rate
of the Ly$\alpha$ selected objects is flat, or possibly rising through
this redshift range, with a value greater than 0.01 $M_{\odot}$ Mpc$^{-3}$
for $q_0 = 0.5$ and $H_0 = 65\ {\rm km}\ {\rm s}^{-1}\ {\rm
Mpc}^{-1}$.  At the highest redshifts, most of the star formation may be
occurring in objects of this class.

\section{Spectroscopic Confirmation of Candidate Lyman alpha emitters}

Paper I\markcite{smitty1} provides a sample of 12 objects in the
SSA22 and HDF fields, with $W_{\lambda} > 100$
\AA\ in the 5390/77 \AA\ filter.  A further 7 objects were detected in a
similar field surrounding the $z\sim 4.53$ quasar BR0019--1522
(Storrie-Lombardi et al.\markcite{sto96} 1996), giving a total sample of
19 objects over an area of $75\,\sq'$.  All the SSA22 and HDF candidate
objects have now been spectroscopically observed using LRIS on the Keck II
telescope, as have three of the objects in the BR0019--1522 field.
Multi-slit masks with $1\farcs4$ slits were used, giving a resolution of
16~\AA\ (Cowie et al.\markcite{large_sample} 1996), and exposure times
ranged from $1-3$ hrs, with most objects having the longer exposure.
Two objects in the HDF field have strong \civ\ emission, placing them at
$z=3.40$ but suggesting AGN activity (Fig.~\ref{fig:1}). These objects
appear similar to the brighter objects in Pascarelle et al.\markcite{pasc96} 
(1996).  However, the remaining objects have only one strong emission
line, and the continuua are too weak for absorption line identifications.

\vspace{0.15in}
%
%
\vspace*{-1.3cm}
\hbox{~}
\centerline{\psfig{file=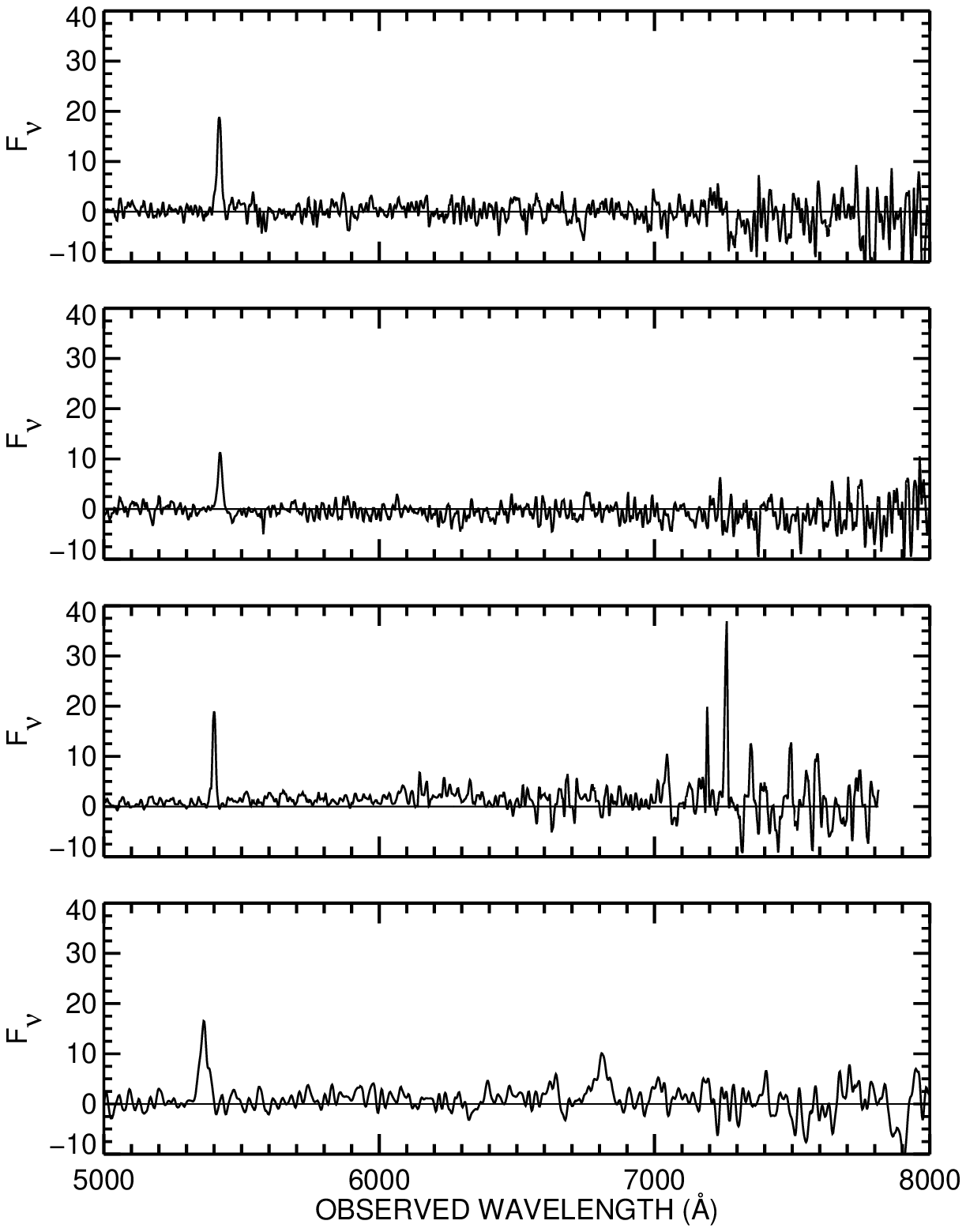,angle=0,width=3.3in}}
\vspace*{-0.7truecm}
\noindent{\scriptsize
\addtolength{\baselineskip}{-3pt}
\figurenum{1}
Figure~1.
Spectra of strong Ly$\alpha$ emitters in SSA22, LA1 and LA3,
({\it top two panels}) compared with the spectrum of an
[O\,{\sc{ii}}] emitter with extremely high equivalent width ({\it third
panel}) shifted so the [O\,{\sc{ii}}] emission lies at the narrowband
wavelength (5390 \AA).  We also show for comparison one of the two Ly$\alpha$
emitters (LA1) in the HDF field that shows strong \civ\ emission ({\it
bottom panel}). The observed equivalent widths of the two SSA22 objects
are $325$ \AA\ for the topmost object and $>845$ \AA\ for the second.
The observed equivalent width of the [O\,{\sc{ii}}] line in the
comparison object is 220 \AA.  The H$\beta$+[O\,{\sc{iii}}] complex
is clearly evident in the case of the [O\,{\sc{ii}}] emitter, and these
cases may be easily distinguished from genuine high-$z$ Ly$\alpha$
emission objects.\label{fig:1}

\addtolength{\baselineskip}{3pt}
}

A primary concern in the simple equivalent width selection procedure is
whether we are misinterpreting emission lines such as
\oii\ 3727~\AA\ and H$\alpha$ 6563~\AA, that are common features in
galaxies at moderate redshift, as higher redshift Ly$\alpha$ emission.
For objects where the continuum is strongly detected it is possible to
use color information to diagnose the different lines (Paper
I\markcite{smitty1}), but in cases in which only one emission line is
seen, we can only make the distinction based on the equivalent width
strength of the line, assuming that an observed equivalent width
$W_{\lambda}\gg100$ \AA\ is generally Ly$\alpha$ rather than unusually
strong \oii\ or H$\alpha$.  Spectroscopic followup is therefore
essential to decide whether this procedure is reliable.

It is easiest to argue that single-line objects are Ly$\alpha$ emitters
at lower redshifts, where the number of possible contaminating lines is
smaller.  An emission line at 5390 \AA\ lies below rest-frame
H$\alpha$, and only slightly above the \oiii+H$\beta$ complex at 5000
\AA, so that the only probable sources of confusion are \oii\ 3727
emitters at $z=0.45$.  However, such objects would then have 
H$\beta$ 4861 and \oiii\ 5007 lying at 7030 \AA\ and 7241 \AA, and we
would expect these emission lines to be strong, given the necessary
strength of the \oii\ 3727 line.  In Figure~\ref{fig:1} we compare the
spectra of two strong emitters ($W_{\lambda}=325$ and $>845$ \AA)
found in the SSA22 field with one of the strongest equivalent width
\oii\ emitters ($W_{\lambda} = 220$ \AA) found in the Hawaii Deep Field
Survey of Cowie et al.\markcite{large_sample} (1996), and with one of
the objects with Ly$\alpha$ and \civ\ emission detected in the HDF.
The \oii\ comparison spectrum has been wavelength shifted so that its
\oii\ line lies over the emission lines in the objects.  It can be
clearly seen that the absence of the H$\beta$+\oiii\ complex in the
SSA22 spectra rules out the possibility that the objects are
\oii\ emitters, and leaves us only with the Ly$\alpha$ interpretation.
\vskip0.25in
%
%
\vspace*{-1.3cm}
\hbox{~}
\centerline{\psfig{file=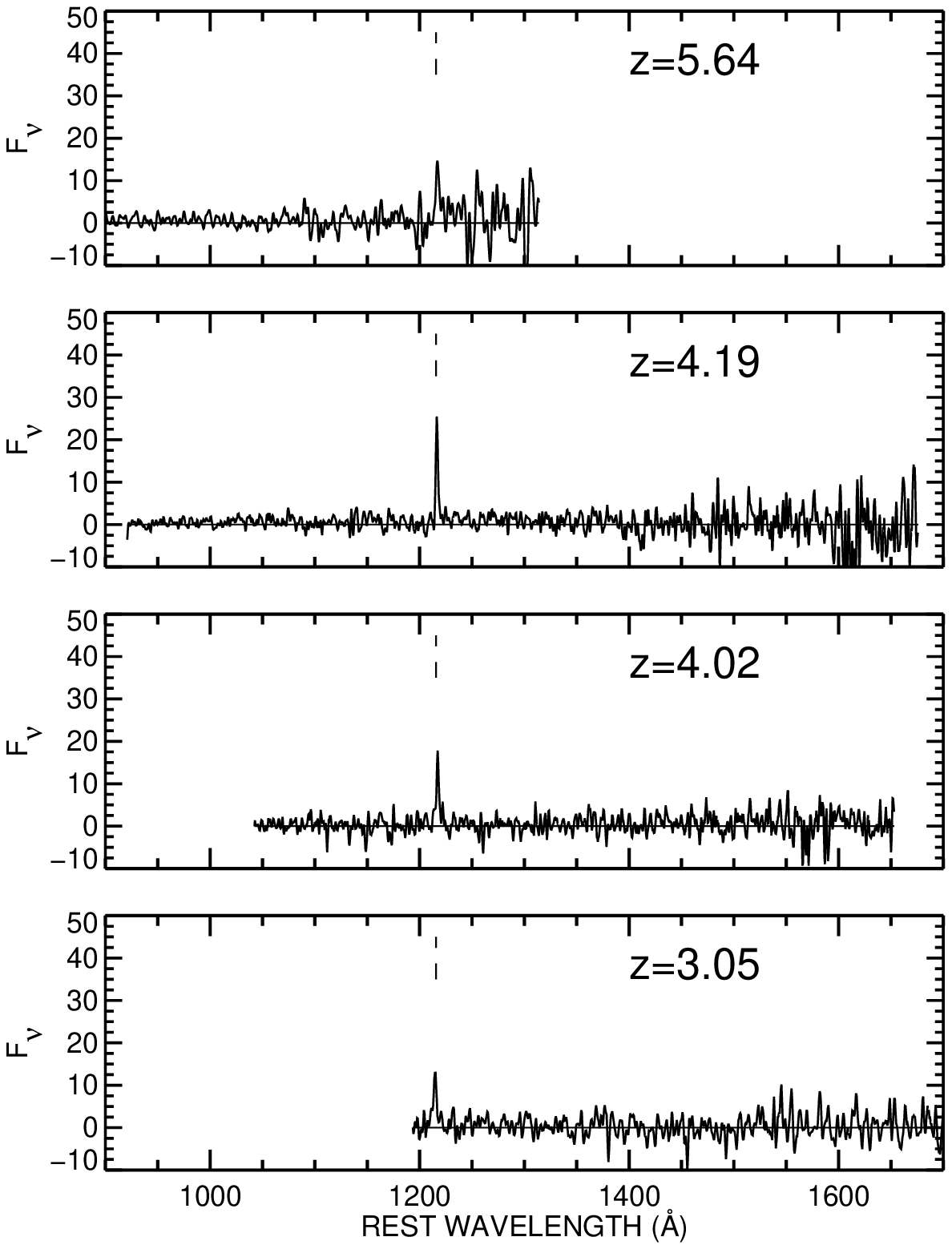,angle=0,width=3.0in}}
\vspace*{-0.7truecm}
\noindent{\scriptsize
\addtolength{\baselineskip}{-3pt}
\figurenum{4}
Figure~4.
The plots show Ly$\alpha$ emission-line objects recovered from deep
multi-slit spectroscopy of a field around the quasar BR1202--0725 using
LRIS.  The 4 detected objects have observed equivalent widths of
$>600$, $>600$, 450, and 350 \AA. See Fig.~3 for the corresponding
two-dimensional spectra. If the emission line were [O\,{\sc{ii}}], the
H$\beta$+[O\,{\sc{iii}}] complex would lie in the $1600-1630$
\AA\ region.\label{fig:4}

\addtolength{\baselineskip}{3pt}
}

At higher redshifts, the most problematic contaminants are likely to be
high ionization extragalactic H{\sc{ii}} regions, with strong lines
of \oiii\ and H$\alpha$.  Spectroscopic followups of candidate emitters
from narrowband surveys where the filter bandpass lies in regions
devoid of strong night sky lines, as is the case for the 5390 \AA\ and
6741 \AA\ filters, can rule out \oiii\ by the doublet line structure.
H$\alpha$ from high ionization systems can have observed equivalent widths 
of several hundred angstroms (cf., Stockton \& Ridgway{\markcite{stock}}
1998), and requires broad wavelength coverage of the \oiii\ lines and/or
the measurement of a continuum break across the emission line to confirm
it as Ly$\alpha$.  We do not find such contaminants evident among the
narrowband and spectroscopic samples discussed in the present paper.

\section{Lyman alpha emitters at $z=4.52$}

Narrowband observations of the SSA22 field were obtained on the
nights of UT 1997 September 21--22 with the LRIS
camera on Keck II, through a narrowband interference filter centered at
6741 \AA, with a bandpass of 78 \AA\ and transmission of 81\%.
The net exposure of 2.7 hrs was
obtained as a sequence of 900 s exposures taken with $10''$ offsets,
with FWHMs of $0\farcs85$ (1.2 hrs) and $0\farcs65$ (1.5 hrs). A
corresponding 1.2 hr $R$-band image was obtained as a series of 360
s exposures to define the local continuum.  Three $W_{\lambda} > 100$
\AA\ objects were detected, which are marked on the narrowband image
in Figure~\ref{fig:2} (Plate 1), and shown as 2-dimensional spectra
in Figure~\ref{fig:3} (Plate 2). 

At this narrowband wavelength (6741 \AA) we may be contaminated by
\oii\ 3727 at $z=0.8$, H$\alpha$ at $z=0.03$, or \oiii\ or H$\beta$ at
$z\sim0.37$.  However, our spectroscopic followup again easily
distinguishes such cases.  The lowest flux object, with the weakest
equivalent width constraint ($W_{\lambda} = 120$ \AA), is identified as
an \oii\ emitter at $z=0.814$, but the stronger object with substantial
red wavelength coverage does not have
detected \oiii\ and does not permit this interpretation.  For the second
strong object there is a strong break across the emission line, consistent
with the Ly$\alpha$ interpretation.  Nor does either
of the objects have emission lines near the primary line, that would
permit the objects to be identified as \oiii\ or H$\beta$.  Once again
there appears no choice but to interpret these lines as Ly$\alpha$.
\vskip0.4in
%
%
\vspace*{-1.3cm}
\hbox{~}
\centerline{\psfig{file=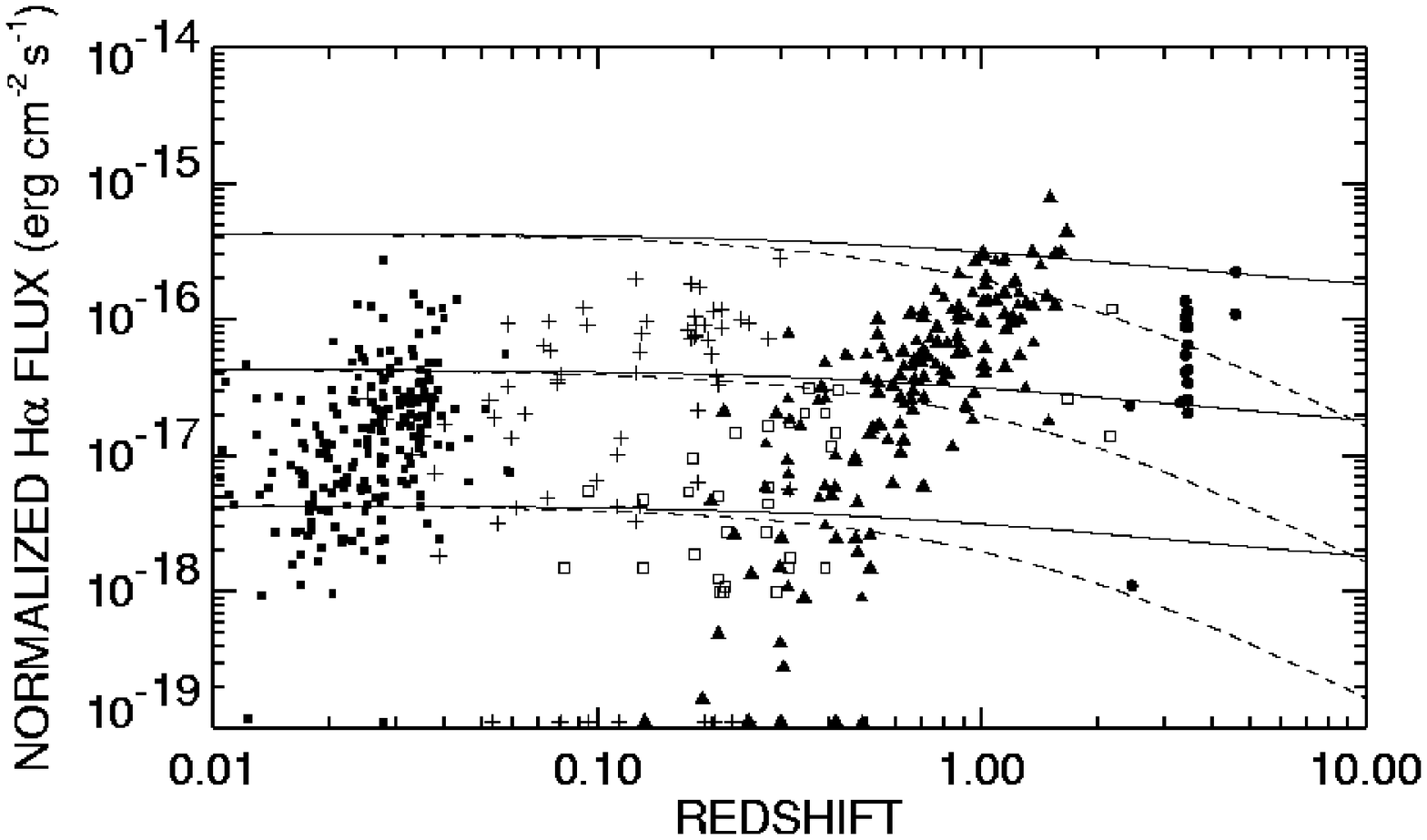,angle=0,width=3.3in}}
\vspace*{-0.7truecm}
\noindent{\scriptsize
\addtolength{\baselineskip}{-3pt}
\figurenum{5}
Figure~5.
Composite plot of H$\alpha$ flux times redshift squared versus
redshift.  At low redshift we show the $f({\rm H}\alpha$+[{\rm
N}\,{\sc{ii}}])$\times{z}^2$ fluxes ({\it filled squares\/}) from the $z
\la 0.045$ sample of Gallego et al.\ (1995),
converted to corresponding H$\alpha$ flux by multiplying by 0.75.
The crosses show normalized H$\alpha$ fluxes or converted, normalized
[O\,{\sc{ii}}] fluxes from the Hawaii wide-field surveys (Songaila et
al.\ 1994).  The filled triangles show renormalized, converted
[O\,{\sc{ii}}] fluxes from a $B=25$ spectroscopic sample (cf., Cowie et
al.\ 1997) with open squares showing renormalized
$f({\rm H}\alpha$+[{\rm N}\,{\sc{ii}}])$\times{z}^2$ or with circles
showing renormalized $f$(Ly$\alpha$) at the highest redshifts if the
spectrum does not cover [O\,{\sc{ii}}].  Finally, the remaining
circles (at $z\sim3.43$ and $z\sim4.54$ show $f({\rm Ly}\alpha) \times
z^2/8.7$
for the samples of objects detected in the two narrowband filters.
The solid ($q_0 = 0.5$) and dashed
($q_0 = 0.02$) reference lines show the fluxes corresponding to star
formation rates $\dot{M}$ = 10, 1, and 0.1 $M_{\odot}$ yr$^{-1}$ for
$H_0=65$ km s$^{-1}$ Mpc$^{-1}$ and the
$\dot{M}$ normalization discussed in the text.\label{fig:5}

\addtolength{\baselineskip}{3pt}
}

\section{Spectroscopic Search}

The surface density of emitters from the narrowband search is
sufficiently high ($\sim15,000$/$\sq^{\circ}$/unit $z$ at $z=3.4$, based
on the 3 fields) that searches of existing very deep long-slit or
multi-slit LRIS observations should yield such objects over the optical
redshift range ($z\sim2 - 7$).  Indeed, Franx et al.\markcite{franx}
1997 identified a second $z\sim 4.92$ object by inferring that the
single strong emission line seen in a serendipitously placed slit
during spectrosopic exposures of their color--selected, gravitationally
lensed galaxy, was also due to Ly$\alpha$ at their target object's
redshift.  In our own archival LRIS data the best suited exposure is
that around the quasar BR1202--0725, where a 6 hr integration in a
single mask configuration was obtained under photometric conditions
and with reasonable seeing (UT 31 March -- 2 April 1997; 4 hrs
obtained with $0\farcs8-0\farcs85$ FWHM, and the remaining 2 hrs with
$\sim0\farcs95$ FWHM).  The data were taken to study color--selected
and emission-line--selected objects in the quasar field 
(Hu, McMahon, \& Egami\markcite{br1202} 1996; Giallongo et 
al.\markcite{br1202_color} 1998) and the mask comprised only twelve slits, 
with the blank sky area covering roughly 200 arcsec$^2$.  A search of this
spectral image yielded four extremely high equivalent width emitters
ranging in redshift from $z=3.04 - 5.64$.  These spectra are shown in
Fig.~\ref{fig:4} and in Plate 2. The two highest redshift objects 
can be seen in
the red continuum, and have a strong break across the emission
line, consistent with the expected Ly$\alpha$ forest decrement.

The surface density of emitters in the spectroscopic field is roughly
60,000 deg$^{-2}$/unit $z$ in the accessible redshift range, consistent
with the narrowband searches when account is made for the slightly
fainter flux limit in the spectrocopic data ($1\ \sigma \sim 10^{-18}$
ergs cm$^{-2}$ s$^{-1}$) and the current uncertainty in the numbers.
Ultimately, the census of the populations is best done with the
narrowband data with its uniform selection procedures and well-defined
flux limits, but the present data suggest that we will see substantial
populations of these objects to very high redshift.

\section{Discussion}

Since resonant scattering enhances the effects of extinction, it is
harder to convert the Ly$\alpha$ emission into a massive star formation
rate than it is for line luminosity diagnostics such as H$\alpha$ and
\oii\ 3727.  For the present calculation, we assume that extinction may
be neglected in computing the required massive star
formation rates, which then constitute a minimum estimate.
However, upward corrections to this value are unlikely to be larger
than a factor of two, since the observed rest-frame equivalent widths
lie in the $100-200$ \AA\ range --- close to the maximum values that
are obtainable from ionization by a massive star population (Charlot \&
Fall\markcite{charl93} 1993).  Then, assuming case B recombination, we
have $L($Ly$\alpha$) = 8.7\,$L($H$\alpha$) (Brocklehurst\markcite{brock}
1971), which using Kennicutt's\markcite{kenn83} (1983) translation of
$\dot{M}$ from H$\alpha$ luminosity, gives
$\dot{M}=(L($Ly$\alpha$)/$10^{42}$ ergs s$^{-1}$) $M_{\odot}$
yr$^{-1}$.  In order to cross-calibrate to \oii\ fluxes at lower
redshift we assume $f$(H$\alpha$+\nii) = $1.25\,f$(\oii) based on the
mean values of the ratio in both the Gallego et al.\markcite{gal95}
(1995) and Hawaii Deep Survey (Cowie et al.\markcite{large_sample}
1996) samples.  We also assume $f$(H$\alpha$+\nii) =
$1.33\,f$(H$\alpha$) (Kennicutt\markcite{kenn83} 1983).

In Figure~\ref{fig:5} we compare the range of Ly$\alpha$ luminosities to the
range of line luminosities in lower redshift objects.  The plot shows the
quantity $z^2\,f$ vs redshift, where H$\alpha$+\nii, \oii, and Ly$\alpha$
fluxes have been converted to H$\alpha$ fluxes using the relationships above,
and we have restricted ourselves to the imaging data in which the fluxes of
the Ly$\alpha$ are well determined.  The comparison objects are drawn
from the surveys of Gallego et al.\markcite{gal_95} 1995 ({\it filled
boxes}), Songaila et al.\markcite{ksurvey_3} 1994 ({\it pluses}), and the
Hawaii $B=25$ sample [{\it open boxes\/} (H$\alpha$+\nii), {\it 
triangles\/} (\oii), and {\it circles\/} (Ly$\alpha$)]. 
The solid ($q_0=0.5$) and dashed
($q_0=0.02$) lines on Fig.~\ref{fig:5} show the fluxes corresponding to
stellar mass production rates of 10 $M_{\odot}$ yr$^{-1}$, 1 $M_{\odot}$
yr$^{-1}$, and 0.1 $M_{\odot}$ yr$^{-1}$ for $H_0 = 65\ {\rm km}\ {\rm
s}^{-1}\ {\rm Mpc}^{-1}$.  Maximum formation rates locally are around a
few $M_{\odot}$ yr$^{-1}$, rising to values of just over 10 $M_{\odot}$
yr$^{-1}$ above $z=0.6$.  The Ly$\alpha$ fluxes at the higher redshifts
are then consistent with this value or just slightly smaller depending on
the extinction correction.  The CADIS results (Thommes et al.\markcite{cadis}
1998) would lie a factor of several times higher in flux than the two Hawaii 
filter samples, but both Keck spectroscopy and repeat Fabry-P\'erot
observations have disproved the original $z\sim5.7$ candidate selection 
(Meisenheimer\markcite{meisen} 1998).

The minimum integrated star formation rates at $z=3.4$ and $z=4.5$ are 0.006
$M_{\odot}$ Mpc$^{-3}$ yr$^{-1}$ and 0.01 $M_{\odot}$ Mpc$^{-3}$
yr$^{-1}$ respectively for $q_0=0.5$ where the first value is slightly
smaller than that given in Paper I since AGN-like objects are
excluded.  Both values are lower limits calculated in the absence of extinction
and the $z=4.5$ value is based on a single field.  
Within the substantial uncertainties of the as yet small 
samples, the results suggest that the star
formation rates in the strong emission line population are constant 
or may possibly be increasing with redshift from
$z=3 - 6$, in contrast to color-based samples where the rate is
declining at higher redshifts (Madau et al.\markcite{madau96} 1996,
\markcite{madau98} 1998).  This is consistent with the
broad expectation that as we move to higher redshifts and earlier
stages of galaxy formation, an increasingly larger fraction of the
star formation should be in strong Ly$\alpha$ emitters that correspond
to the youngest galaxies.

\acknowledgements

We thank T.\ Bida, R.\ Goodrich, J.\ Aycock, R.\ Quick, and W.\ Wack for
their assistance in obtaining the observations, which were made
possible by the LRIS spectrograph of Judy Cohen and Bev Oke. We would
also like to acknowledge helpful communications from K. Meisenheimer.
This work was supported by the State of Hawaii and by NASA grants
GO-5975.01-94A, GO-5922.01-94A, GO-6626.01-95A, and AR-6337.06-94A from
Space Telescope Science Institute, which is operated by AURA, Inc.,
under NASA contract NAS 5-26555.  E.M.H.\ acknowledges a University
Research Council Seed Money grant. R.G.M. thanks the Royal Society for
support.

\newpage
\begin{landscape}
%
%
\raggedbottom
\addtolength{\topskip}{0pt plus .0001fil}
\renewcommand{\newpage}{\par \break}
\begin{figure}[h]
\figurenum{2}
\centerline{\psfig{file=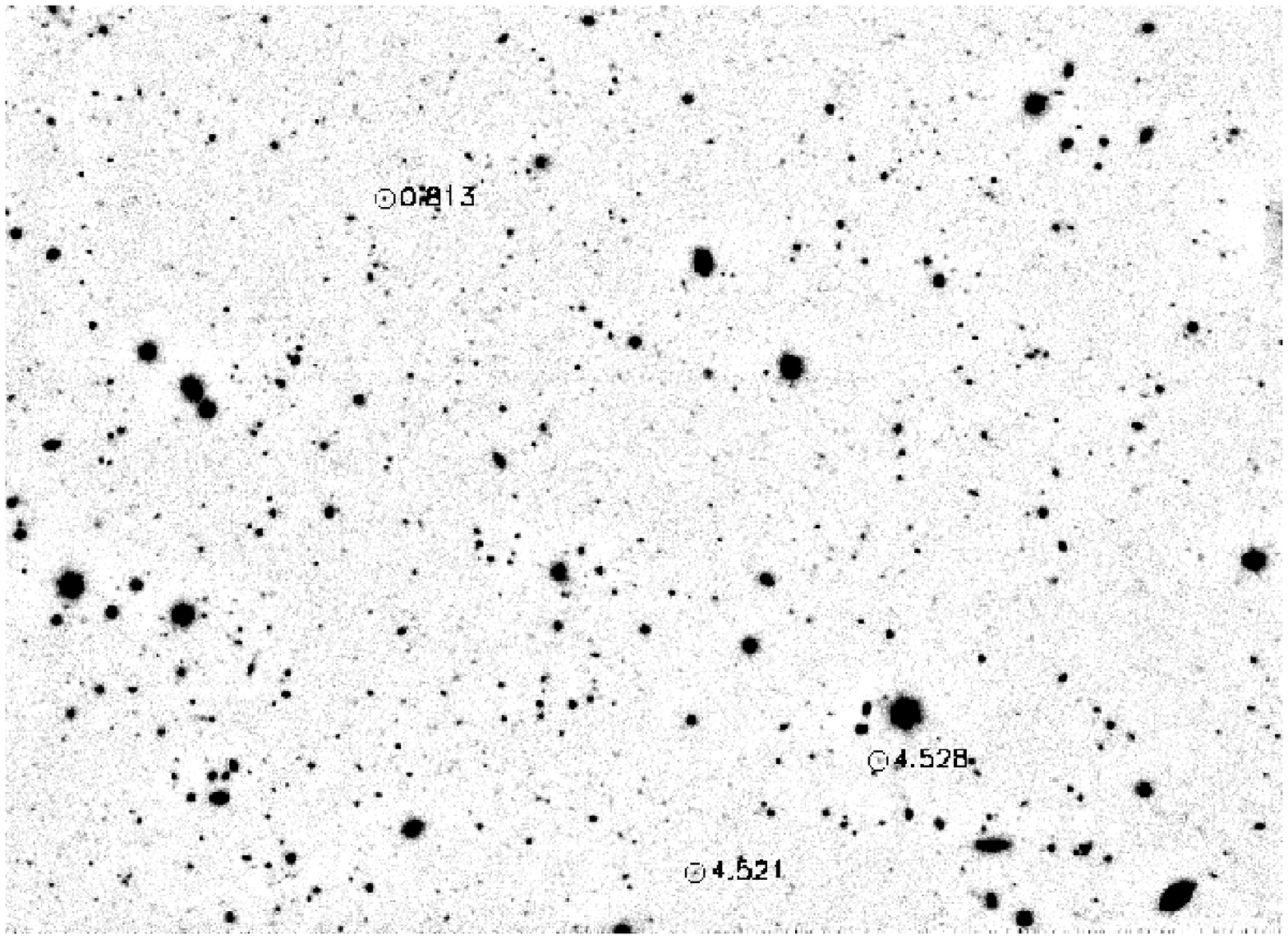,angle=0,width=8.2in}}
\caption{(Plate 1) A 2.7 hr Keck narrowband image of the $6' \times
4'$ region around the SSA22 field through a 78 \AA\ filter centered at
6741 \AA.  Objects with observed equivalent widths $W_{\lambda} > 100$
\AA\ are circled.  The object with the lowest observed equivalent width
($W_{\lambda}=120$ \AA) is spectroscopically identified as a $z=0.813$
[O\,{\sc{ii}}] emitter.  However, the two higher equivalent width
objects appear to be Ly$\alpha$ emitters ($W_{\lambda}=200$ \AA,
$z=4.521$ and $W_{\lambda}=1000$ \AA, $z=4.528$).}\label{fig:2}
\end{figure}

\newpage
\begin{figure}
\figurenum{3}
~\vskip-0.75in
\centerline{\psfig{file=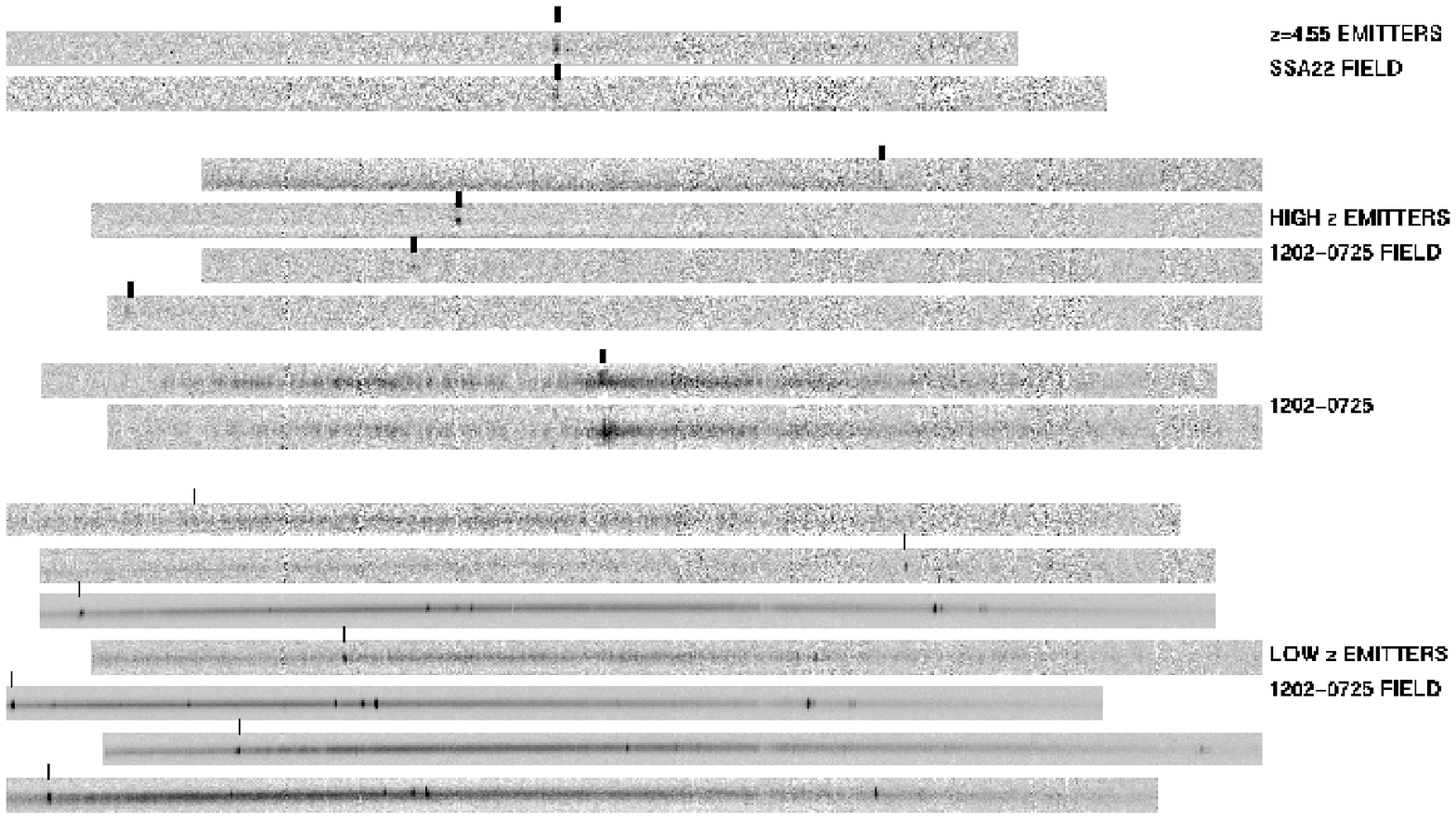,angle=0,width=8.2in}}
\caption{(Plate 2) Gray scale images of emission-line objects selected
in the spectroscopic observation of the BR1202--0725 field and of
the two emitters in the SSA22 field (Plate 1) selected in the
6741/78 \AA\ filter ({\it top two panels\/}).  The images have been
registered in wavelength with a range from 4400 \AA\ to 9700 \AA.
The lower seven panels show emission-line objects at $z<1.20$ in
the spectral image.  For each object, the \oii\ $\lambda\,3727$ line
is indicated.   The two middle panels show the $z=4.69$ quasar,
BR1202--0725, observed at two different orientations.  Narrow, spatially
separated Ly$\alpha$ emission without strong continuum is seen at both
angles, corresponding to the two companions to the quasar (Hu, McMahon, \&
Egami 1996; Fontana et al.\ 1996; Petitjean et al.\ 1996; Hu, McMahon, \&
Egami 1997).  The next four panels show the single emission-line objects
({\it thick tick marks\/}) in the BR1202--0725 field.  The two longer
wavelength objects at $z=4.19$ and $z=5.64$ have continuum breaks across the
emission line of $2.3\{^{2.6}_{2.0}$ and $2.6\{^{5.2}_{1.8}$ respectively,
suggesting that this line is Ly$\alpha$, while the two lower wavelength
lines, where no continuum is visible, have none of the additional emission
lines which would be seen if they were at lower redshift.  (The longest
wavelength emission line object lies close to a bright foreground galaxy at
$z=0.236$, whose spectrum can be seen at the bottom edge of the image.
However, the required velocity separation of $-1500$ km s$^{-1}$ would seem
large for the line to be interpreted as H$\alpha$ from an extragalactic
H{\sc{ii}} region associated with this object.)\ \ For the SSA22
($z\sim4.52$) objects, the spectrum in the upper panel shows a continuum
break, while the lower panel's spectrum is only an isolated emission
line.}\label{fig:3}
\end{figure}
\end{landscape}
\end{document}